\documentclass[10pt,twocolumn,amsmath,amssymb,aps,prb,showpacs,longbibliography,superscriptaddress]{revtex4-1}
\usepackage[latin9]{inputenc}
\usepackage{mathrsfs}
\usepackage{amsmath}
\usepackage{amssymb}
\usepackage{graphicx}
\usepackage{esint}

\makeatletter

\usepackage{amsfonts}
\usepackage{graphics}

\makeatother

\begin{document}
\title{Fulde-Ferrell-Larkin-Ovchinnikov States and Topological Bogoliubov
Fermi Surfaces in Altermagnets: an Analytical Study}
\author{Zhao Liu}
\affiliation{Centre for Quantum Technology Theory, Swinburne University of Technology,
Melbourne 3122, Australia}
\author{Hui Hu}
\affiliation{Centre for Quantum Technology Theory, Swinburne University of Technology,
Melbourne 3122, Australia}
\author{Xia-Ji Liu}
\affiliation{Centre for Quantum Technology Theory, Swinburne University of Technology,
Melbourne 3122, Australia}
\date{\today}
\begin{abstract}
We present an analytical study of the ground-state phase diagram for
dilute two-dimensional spin-1/2 Fermi gases exhibiting $d$-wave altermagnetic
spin splitting under $s$-wave pairing. Within the Bogoliubov-de Gennes
mean-field framework, four distinct phases are identified: a Bardeen-Schrieffer-Cooper-type
superfluid, a normal metallic phase, a nodal superfluid with topological
Bogoliubov Fermi surfaces (TBFSs), and Fulde-Ferrell-Larkin-Ovchinnikov
(FFLO) states with finite center-of-mass momentum. Among these, the
FFLO states and TBFSs exemplify two unconventional forms of superconductivity.
Considering the simplicity of this model, with only one band, zero
net magnetization, and $s$-wave paring, the emergence of both unconventional
phases underscores the pivotal role of altermagnetic spin splitting
in enabling exotic pairing phenomena. This analytical study not only
offers a valuable benchmark for future numerical simulations, but
also provides a concrete experimental roadmap for realizing FFLO states
and TBFSs in altermagnets. 
\end{abstract}
\maketitle

\section{Introduction}

The Fulde-Ferrell-Larkin-Ovchinnikov (FFLO) state \citep{Fulde1964, Larkin1964}
was first proposed in the 1960s in the context of Bardeen-Schrieffer-Cooper
(BCS) superconductivity \citep{BCS1957} under strong spin-exchange
fields. In the presence of spin-population imbalance, Cooper pairs
acquire a finite center-of-mass momentum ${\bf q}$ to minimize the
free energy, therefore the translational symmetry of the
homogeneous superfluidity (SF) is broken. Over the past six decades, extensive
efforts have been devoted to realizing the FFLO phase across a wide
range of platforms, including high-density quark matter \citep{Casalbuoni2004, Alford2008},
ultracold Fermi gases \cite{Hu2006,Hu2007,Liu2007,Radzihovsky2009,Liao2010,Liu2013,Cao2014,Takahashi2014,Sheehy2015,Wang2018,Kawamura2022},
and electronic materials \citep{Shimahara1994, Pickett1999, Radovan2003, Uji2006, Matsuda2007, Kenzelmann2008, Mayaffre2014, Cho2017, Liu2017, Zhao2023, Wan2023}.
However, conclusive evidence-namely, spatial modulation of either
the phase or amplitude of the SF order parameter-remains elusive.

Altermagnetism (AM) \citep{Smejkal2022-1}, a novel type of collinear
magnetism beyond the traditional ferromagnetism (FM) and antiferromagnetism
(AFM), has recently attracted considerable attention \citep{Smejkal2022-2, Mazin2022, Bai2024, Song2025, Guo2025, Jungwirth2025-1, Jungwirth2025-2}.
As a hybrid of FM and AFM, AM features compensated collinear spin
configurations in real space (akin to conventional AFM) while exhibits
non-relativistic spin splitting (NRSS) in momentum space (reminiscent
of FM) \citep{Lee2024, Souma2024, Reimers2024, Ding2024, Krempasky2024, Yang2025, Leiviska2024, Han2024,  Zhou2025, Jiang2025, Zhang2025}.
Combing both merit of FM and AFM, AM offers exciting opportunities
for novel applications in spintronics \citep{Naka2019, Smejkal2020, Reichlova2024},
quantum geometry \citep{Fang2024, Gong2024, LiY2024-1, Hernandez2025, Liu2025, Antonenko2025},
and superconductivity \citep{Mazin2022-2, Zhu2023, Brekke2023, Bose2024, Wu2025, Maeda2025, LiY2024-2,  Chakraborty2024-2, LiY2023, Ghorashi2024, Mondal2025, Hodge2025, Ouassou2023, Lu2024, Fukaya2025, Zhao2025, Banerjee2024, Chakraborty2024-3, Papaj2023, Sun2023, Das2024, Wei2024, Carvalho2024, Parthenios2025, Fukaya2025-JPCM, Liu2025-2}.

AM also provides a promising platform for the FFLO state for two main
reasons. First, FFLO phases typically require a nonzero spin imbalance,
which is usually achieved via external Zeeman fields. However, in
conventional superconductors, the orbital effect induced by magnetic
fields can suppress the superconductivity before the FFLO emerges
\citep{Maki1966, Gruenberg1966}. In contrast, AM naturally hosts
NRSS without the need of external magnetic field, thereby avoiding
detrimental orbital effects. Second, conventional FFLO state involves
spontaneously breaking of the continuous $U(1)$ rotational symmetry
of the Fermi surfaces, which, according to Nambu-Goldstone theorem
\citep{Nambu1960, Goldstone1961, Goldstone1962}, leads to gapless
modes associated with fluctuations between equivalent ${\bf q}$.
In AM, the continuous rotational symmetry is reduced to a discrete
one \cite{Wu2007}, significantly suppressing such fluctuations. These
unique advantages have motivated several theoretical investigations
of FFLO phases in altermagnetic metals \citep{Feiguin2009, Soto-Garrido2014, Gukelberger2014, Sumita2023,  Chakraborty2024-1, ZhangS2024, Hong2025, Hu2025-1, Hu2025-2, Sim2025, Sumita2025}.

Despite these efforts, the existence of a  FFLO state in a two-dimensional
(2D) $d$-wave AM Fermi gas with purely $s$-wave pairing remains controversial. 
Numerical simulations on a 2D square lattice by Feiguin \textit{et al.} \citep{Feiguin2009}, Gukelberger \textit{et al.} \citep{Gukelberger2014}, and Chakraborty \textit{et al.} \citep{Chakraborty2024-1} reported no evidence of a FFLO state within the parameter regimes explored. This  absence was further supported by the analytical analysis of the corresponding continuum model according to 
Soto-Garrido \textit{et al.} \citep{Soto-Garrido2014}.
However, more recent studies
challenge this conclusion. Using a lattice model closely related to that of  Chakraborty \textit{et al.} \citep{Chakraborty2024-1}, Hong \textit{et al.} \citep{Hong2025} found that a FFLO state does emerge under
purely $s$-wave pairing. Consistent with this, Hu \textit{et al.} \citep{Hu2025-1}
conducted a numerical calculation of the continuum model close to that of 
Soto-Garrido \textit{et al.} \citep{Soto-Garrido2014} and confirmed
the presence of a FFLO state. These conflicting results raise a fundamental
question: \textit{Can a 2D $d$-wave AM Fermi gas support a FFLO state driven
solely by $s$-wave pairing?} 

\begin{figure}
\begin{centering}
\includegraphics[width=0.5\textwidth]{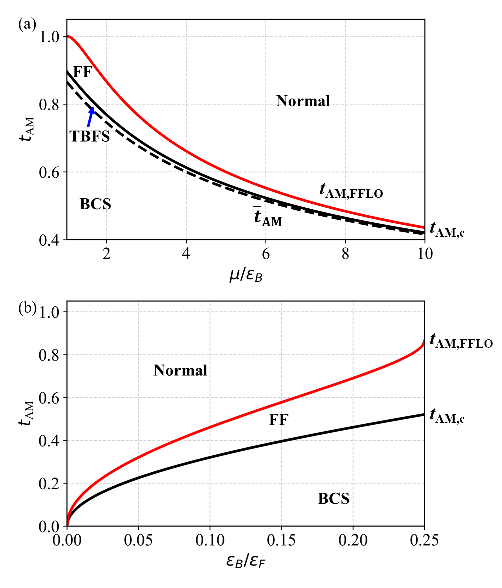}
\par\end{centering}
\caption{(a) Zero temperature phase diagram at fixed chemical potential $\mu$ (normalized to the two-body binding energy $\varepsilon_{B}$) versus non-relativistic spin splitting $t_{\textrm{AM}}$. The solid black/dashed black/solid red curve giving the phase boundary is determined by Eq. (\ref{eq: t_am_c-mu})/Eq. (\ref{eq: upper-t_AM-0-mu})/Eq. (\ref{eq: t_AM-FFLO-mu}), respectively.
(b) Zero temperature phase diagram at fixed total particle number $N$ as a function of the binding energy $\varepsilon_{B}$ normalized to the Fermi energy $\varepsilon_{F}$. The solid black/solid red curve is given by Eq. (\ref{eq: t_am_c-N})/Eq. (\ref{eq: t_AM-FFLO-N}), respectively. 
The black curves in both figures show the Pauli-limit, upper altermagnetic coupling strength, where a SF with zero center-of-mass momentum $\mathbf{q}=0$ becomes unstable towards a normal state, when the possibility of an inhomogeneous FFLO is not considered. In the weak-coupling regime (i.e., for large $\mu/\varepsilon_{B}$ or small $\varepsilon_{B}/\varepsilon_{F}$), the solid black curve in (a) and (b) provides an excellent estimate for the phase boundary separating a TBFS phase and a FFLO state at fixed $\mu$, and a BCS SF phase and a FFLO state at fixed $N$, respectively. }
\label{fig: Fig-1} 
\end{figure}

\subsection{Summary of main results}
In this work, we give a definitive and affirmative answer by \textit{analytically}
determining the ground-state phase diagram at $T$ = 0 K. We consider two physically relevant settings: (i) fixed $\mu/\varepsilon_{B}$, where $\mu$
is the total chemical potential and $\varepsilon_{B}$ is the two-body
binding energy which is experimentally controllable in the ultracold
atom systems; and (ii) fixed total particle number $N$, which is 
appropriate for describing electronic superconductors. 

Fig.\ref{fig: Fig-1}(a) shows the phase diagram in the fixed-chemical-potential ensemble. We find a finite window of FFLO stability that expands with increasing $d$-wave altermagnetic spin splitting $t_{AM}$ (a dimensionless parameter defined below). Adjacent to the FFLO state is a nodal SF hosting topological Bogoliubov Fermi surfaces (TBFSs), whose region of stability is comparatively narrow.  

Fig.\ref{fig: Fig-1}(b) shows the phase diagram at fixed total particle number. Here, the horizontal axes is normalized to the Fermi energy $\varepsilon_F$ (defined below), which is determined by the imposed particle number. In this ensemble, we again find that the FFLO window broadens with increasing $t_{AM}$, appearing between the conventional BCS SF and the normal altermagnetic metallic phase. 

To obtain the FFLO-normal phase boundary (red curves in Fig.\ref{fig: Fig-1}), one must evaluate the two polar integration integrals $I_3$ and $I_4$ defined in Eq. (\ref{eq: integral-I3-I4}). The central technical contribution of this work is an analytical evaluation of $I_3$ and $I_4$, which, to our best knowledge, has not been carried out exactly in previous works \citep{Soto-Garrido2014, ZhangS2024}. In general, two approaches exist for polar integration in 2D. The first integrates over the radial distance $k$ followed by the polar angle $\theta$. This integration order is widely employed in evaluating 2D polar integration, as $I_1$ and $I_2$ in this work. The second integrates over the polar angle $\theta$ first, which is recast to a complex integral on the unit circle $|z| = 1$ in the complex $z$ plane, and subsequently performs the remaining $k$-integration. This second procedure is used to evaluate $I_3$ and $I_4$. After performing the $\theta$-integration, the resulting expressions depend on both $k$ and the FFLO center-of-mass momentum $\bf{q}$, and acquire different analytic forms in distinct ($k$, $\bf{q}$) domains. The boundaries between these domains determine the critical momentum $\bf{q}_c$ marking the onset of the FFLO state. For $\bf{q}$ oriented along the (1, 1) direction, we found that $|\bf{q}_c|$ admits a simple geometric interpretation: it is the momentum at which the spin-up and spin-down Fermi surfaces become nested (see Fig.\ref{fig: Fig-4}(a)). 

The remainder of the paper is organized as follows. In the next section
(Sec. II), we outline the model Hamiltonian that describes a $d$-wave
altermagnetic metal. In Sec. III, we briefly discuss how to solve
the model Hamiltonian by using the standard Bogoliubov de-Gennes (BdG)
mean-field theory, with an inhomogeneous FF order parameter. In Sec.
IV, we present the phase diagram between normal, nodal SF,
and BCS SF phases. The nontrivial $\mathbb{Z}_{2}$ topology of BFS is characterized
by the Pfaffian of the BdG Hamiltonian. In Sec. V, we discuss in details
the FFLO phase boundary driven by the NRSS. Finally, we conclude in Sec.
VI and present some outlooks for future works.

To simplify the notation, the Fermi wavevector $k_{F}$ and Fermi
energy $\varepsilon_{F}$ are taken as the units for wavevector and
energy, respectively. This means that we can directly set $2m=\hbar=k_{B}=k_{F}=\varepsilon_{F}=1$.

\section{Model Hamiltonian}

\begin{figure*}
\begin{centering}
\includegraphics[width=1\textwidth]{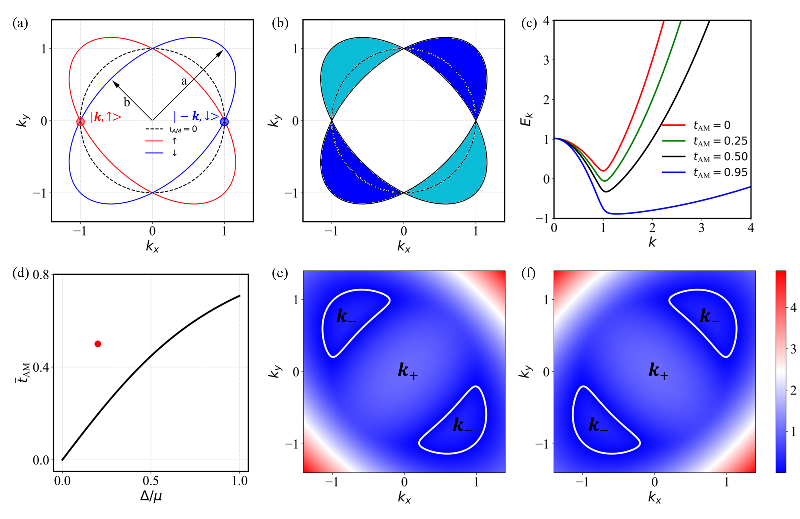} 
\par\end{centering}
\caption{(a) Schematic $d_{xy}$-wave spin splitting obtained by plotting $\xi_{\bf{k}}+s(\sigma)J_{\bf{k}}=0$. Here $\mu=1$. $t_{\textrm{AM}}=0$ gives the dashed circle, and $t_{\textrm{AM}}=0.5$ gives the red and blue ellipses, corresponding to spin-up and spin-down channels, respectively. The semi-major and semi-minor of each ellipse is denoted as $a$ and $b$. 
(b) Contour plot of $E_{\bf{k}\pm}=0$ (see Eq. (\ref{eq: BdG-spectrum}) for definition) at normal phase, here $\mu=1$, $t_{\textrm{AM}}=0.5$, and $\Delta=0$. The dash line corresponds to $\sum_{\eta}E_{\bf{k} \eta}=0$. Cyan and blue colored regions correspond to negative $E_{\bf{k} +}$ and $E_{\bf{k} -}$. 
(c) Plot of $E_{\bf{k}}$ (see Eq. (\ref{eq: BdG-spectrum-min}) for definition) at different $t_{\textrm{AM}}$ values, here $\mu=1,\Delta=0.2$. When $t_{AM}$ is small, $E_{\bf{k}\pm}$ is always positive, indicating a BCS SF phase. When $t_{AM}$ is large, $E_{\bf{k}\pm}$ will vanishing, meaning the emergence of Bogoliubov Fermi surfaces.
(d) Plot of $\bar{t}_{\textrm{AM}}$ (see Eq. (\ref{eq: upper-t_AM-0}) for definition)  as a function of $\Delta/\mu$.
(e)-(f) Anisotropic Bogoliubov quasiparticle energy $E_{\bf{k} +}$
and $E_{\bf{k} -}$ at BCS SF phase. The white solid lines represent the Bogoliubov Fermi surfaces $E_{\bf{k} \pm} = 0$, which separate the whole $\bf{k}$ domain into two regions labelled by $\bf{k}_{\pm}$. Here $\mu=1$, $\Delta=0.2$, and $t_{\textrm{AM}}=0.5$, which corresponds to the red dot in (d). }
\label{fig: Fig-2} 
\end{figure*}

The model Hamiltonian describing two-dimensional spin-1/2 Fermion
gas is: 
\begin{eqnarray}
{\mathscr{H}} & = & \int d{\bf x}({\cal H}_{0}+{\cal H}_{\textrm{int}}),\\
{\cal H}_{0} & = & \sum_{\sigma=\uparrow,\downarrow}\psi_{\sigma}^{\dagger}\left({\bf x}\right)\left[\hat{\xi}_{{\bf k}}+s(\sigma)\hat{J}_{\mathbf{k}}\right]\psi_{\sigma}\left({\bf x}\right),\\
{\cal H}_{\textrm{int}} & = & g\psi_{\uparrow}^{\dagger}\left({\bf x}\right)\psi_{\downarrow}^{\dagger}\left({\bf x}\right)\psi_{\downarrow}\left({\bf x}\right)\psi_{\uparrow}\left({\bf x}\right).
\end{eqnarray}
For the non-interacting part ${\cal H}_{0}$, $\hat{\varepsilon}_{{\bf k}}\equiv-{\bf \nabla}^{2}$
is the operator for kinetic energy where $\hat{k}_{x}=-i\partial_{x}$,
$\hat{k}_{y}=-i\partial_{y}$. Operating on a plane wave with wave vector $\bf{k}$, it gives the kinetic energy $\varepsilon_{\bf{k}} = k^2$, where $k$ is the length of $\bf{k}$. 
$\hat{\xi}_{{\bf k}}\equiv\hat{\varepsilon}_{{\bf k}}-\mu$
is the modified kinetic energy operator where $\mu$ is the chemical
potential. The chemical potential is tuned to yield the given total
particle density $n=n_{\uparrow}+n_{\downarrow}$. $\hat{J}_{\mathbf{k}}\equiv t_{\textrm{AM}}2\hat{k}_{x}\hat{k}_{y}$
where $s(\uparrow)=+1$ and $s(\downarrow)=-1$ is the $d_{xy}$-wave
altermagnetic spin splitting, whose strength is controlled by the
dimensionless parameter $t_{\textrm{AM}}$ (without the loss of generality,
$t_{\textrm{AM}}$ is set to be positive here). In 2D, there are two independent $d$-wave spin splitting: $d_{xy}$ and $d_{x^2-y^2}$. As $d_{x^2-y^2}$ can be obtained from $d_{xy}$ by $\pi/4$ rotation, the results obtained here can be generalized to $d_{xy}$-wave altermagnetic spin splitting straightforwardly. 
Without $\hat{J}_{\mathbf{k}}$,
the spin degenerated Fermi surface is a circle with radius $\sqrt{\mu}$,
which conserves $U(1)$ rotational symmetry (see the dashed circle in
Fig. \ref{fig: Fig-2}(a)). The introduction of $\hat{J}_{\mathbf{k}}$
distorts the Fermi surface into two orthogonal ellipses (denoted as
Fermi ellipse here) with equal semi-major axis $a\equiv\sqrt{\mu/(1-t_{\textrm{AM}})}$
and semi-minor axis $b\equiv\sqrt{\mu/(1+t_{\textrm{AM}})}$ (see
Fig. \ref{fig: Fig-2}(a)), in this regard, $t_{\textrm{AM}}$ also
measures the eccentricity of the Fermi ellipses at a given $\mu$.
These two Fermi ellipses together exhibit $d_{xy}$-wave form (see
Fig. \ref{fig: Fig-2}(a)): spin up and down ellipse has major axis
along $\left(1,-1\right)$ and $\left(1,1\right)$ direction, respectively.
Since the two Fermi ellipses always have the same area even with finite
momentum boost in the FFLO phase, the system remains spin-population balanced in the whole phase diagram. Such a circle-to-ellipse distortion breaks
the continuous $U(1)$ to the discrete $C_{4z}$ symmetry, leaving
four degeneracies along both $(1,0)$ and $(0,1)$ directions. As
mentioned before, such continuous symmetry breaking would stabilize
the FFLO state.

For the inter-particle interaction ${\cal H}_{\textrm{int}}$, we
consider a short-range $s$-wave contact potential. Such a consideration
is straightforward as pair states like $\left|\mathbf{k},\uparrow\right\rangle $
and $\left|-\mathbf{k},\downarrow\right\rangle $ in Fig. \ref{fig: Fig-2}(a)
are susceptible to $s$-wave pairing according to BCS theory \citep{BCS1957}.
In two dimensions, its interaction strength $g$ can be conveniently
regularized by using a two-body binding energy $\varepsilon_{B}$
\citep{He2015}: 
\begin{equation}
\frac{1}{g}=-\frac{1}{\mathcal{A}}\sum_{\mathbf{k}}\frac{1}{2\varepsilon_{\mathbf{k}}+\varepsilon_{B}},\label{eq: nor-cond}
\end{equation}
where $\mathcal{A}$ is the area of the 2D Fermi system.

\section{Mean-field theory}

To search for the FFLO state, for simplicity, we employ a plane-wave ansatz for the pairing order parameter $\Delta({\bf x})=-g\left\langle \psi_{\downarrow}\left({\bf x}\right)\psi_{\uparrow}\left({\bf x}\right)\right\rangle =\Delta e^{i\mathbf{q}\cdot\mathbf{x}}$
at position $\mathbf{x}$ with the center-of-mass momentum $\mathbf{q}$. This is equivalent to the phase-modulated FF form introduced by Fulde and Ferrel \citep{Fulde1964}, here we still refer to the resulting state as the FFLO state. The amplitude-modulated LO type, will be explored in future work. 

Under plane-wave ansatz, $\left|\Delta({\bf x})\right|^{2}=\Delta^{2}$ (below we take $\Delta$
positive), and consider the standard mean-field decoupling of the interaction
Hamiltonian density \citep{Hu2006,Liu2013}, 
\begin{equation}
{\cal H}_{\textrm{int}}\simeq-\left[\Delta({\bf x})\psi_{\uparrow}^{\dagger}\left({\bf x}\right)\psi_{\downarrow}^{\dagger}\left({\bf x}\right)+\text{H.c.}\right]-\frac{1}{g}\Delta^{2}.
\end{equation}

In the framework of BdG mean-field theory, the total Hamiltonian can
be rewritten into the form \citep{Hu2006,Liu2013}, 
\begin{equation}
{\cal \mathscr{H}}_{\textrm{MF}}=\int d{\bf x}\Phi^{\dagger}\left({\bf x}\right){\cal H}_{\textrm{BdG}}\Phi\left({\bf x}\right)-\frac{\Delta^{2}}{g}\mathcal{A}+\mathcal{E}_{0}',
\end{equation}
where $\Phi\left({\bf x}\right)\equiv[\psi_{\uparrow}\left({\bf x}\right),\psi_{\downarrow}^{\dagger}\left({\bf x}\right)]^{T}$
is the Nambu spinor and 
\begin{equation}
{\cal H}_{\textrm{BdG}}\equiv\left[\begin{array}{cc}
\hat{\xi}_{{\bf k}}+\hat{J}_{\mathbf{k}} & -\Delta\left({\bf x}\right)\\
-\Delta^{*}\left({\bf x}\right) & -\left(\hat{\xi}_{{\bf k}}-\hat{J}_{\mathbf{k}}\right)
\end{array}\right],\label{eq: HBdG}
\end{equation}
and $\mathcal{E}_{0}'=\sum_{\mathbf{k}}(\hat{\xi}_{{\bf k}}-\hat{J}_{\mathbf{k}})$
is an energy shift (to be specified below), due to the abnormal order
in the field operators $\psi_{\downarrow}(\mathbf{x})$ and $\psi_{\downarrow}^{\dagger}({\bf x})$
taken in the Nambu spinor representation.

\subsection{Bogoliubov quasiparticles}

As the FF order parameter takes a plane-wave form, it is convenient
to diagonalize the mean-field Hamiltonian in momentum space, with
the following Bogoliubov equations \citep{Liu2013}, 
\begin{equation}
{\cal H}_{\textrm{BdG}}\Phi_{{\bf k}}\left({\bf x}\right)=E_{{\bf k}}\Phi_{{\bf k}}\left({\bf x}\right),
\end{equation}
where 
\begin{equation}
\Phi_{{\bf k}}\left({\bf x}\right)\equiv\left[\begin{array}{c}
u_{{\bf k}\uparrow}e^{+i{\bf q}{\bf x}/2}\\
v_{{\bf k}\downarrow}e^{-i{\bf q}{\bf x}/2}
\end{array}\right]e^{i{\bf kx}}
\end{equation}
and $E_{{\bf k}}$ are the wavefunction and energy of the Bogoliubov
quasiparticles, respectively. With this wavefunction, the energy shift
is $\mathcal{E}_{0}'=\sum_{\mathbf{k}}(\xi_{{\bf k}-\mathbf{q}/2}-J_{\mathbf{k}-\mathbf{q}/2})$,
where $\xi_{\mathbf{k}}\equiv k^{2}-\mu$, and $J_{\mathbf{k}} \equiv t_{\textrm{AM}}2k_{x}k_{y}$.
Then the Bogoliubov equations becomes, 
\begin{equation}
\left[{\cal H}_{\textrm{BdG}}\right]\left[\begin{array}{c}
u_{{\bf k}\uparrow}\\
v_{{\bf k}\downarrow}
\end{array}\right]=E_{{\bf k}}\left[\begin{array}{c}
u_{{\bf k}\uparrow}\\
v_{{\bf k}\downarrow}
\end{array}\right],
\end{equation}
where $\left[{\cal H}_{\textrm{BdG}}\right]$ is a 2 by 2 matrix given
by, 
\begin{equation}
\left[\begin{array}{cc}
\xi_{{\bf k}+\mathbf{q}/2}+J_{\mathbf{k}+\mathbf{q}/2} & -\Delta\\
-\Delta & -(\xi_{{\bf k}-\mathbf{q}/2}-J_{\mathbf{k}-\mathbf{q}/2})
\end{array}\right].
\end{equation}
$\left[{\cal H}_{\textrm{BdG}}\right]$ can be diagonalized with eigenvectors
$[u_{{\bf k}\uparrow}^{(\eta)},v_{{\bf k}\downarrow}^{(\eta)}]^{T}$
and corresponding eigenvalues $\eta E_{{\bf k}\eta}$, labelled by
the branch index $\eta=\pm$. Explicitly, we have the Bogoliubov quasiparticle
energy spectrum, 
\begin{equation}
E_{\mathbf{k}\pm}=\sqrt{A_{\mathbf{k}}^{2}+\Delta^{2}}\pm B_{\mathbf{k}},
\label{eq: BdG-spectrum}
\end{equation}
where 
\begin{eqnarray}
A_{\mathbf{k}} & \equiv & \xi_{{\bf k}}+\frac{{\bf q}^{2}}{4}+t_{\textrm{AM}}(k_{x}q_{y}+k_{y}q_{x}),\\
B_{\mathbf{k}} & \equiv & {\bf k}\cdot{\bf q}+J_{\mathbf{k}}+\frac{t_{\textrm{AM}}}{2}q_{x}q_{y}.
\end{eqnarray}
Here, as usual we have reversed the sign for the eigenvalue $-E_{\mathbf{k}-}$,
in order to write the corresponding field operators of Bogoliubov
quasiparticles $\alpha_{\mathbf{k}-}$ and $\alpha_{\mathbf{k}-}^{\dagger}$
in the normal order. This introduces another energy shift $\mathcal{E}_{0}''=-\sum_{\mathbf{k}}E_{\mathbf{k}-}$
in the mean-field Hamiltonian. When $t_{\textrm{AM}} = 0$,
our quasiparticle energy spectrum reduces to the results of Ref. \citep{Sheehy2015}. In the end, we obtain the
diagonalized mean-field Hamiltonian, 
\begin{equation}
{\cal \mathscr{H}}_{\textrm{MF}}=\sum_{{\bf k},\eta=\pm}E_{{\bf k}\eta}\alpha_{{\bf k\eta}}^{\dagger}\alpha_{{\bf k}\eta}-\frac{\Delta^{2}}{g}\mathcal{A}+\mathcal{E}_{0},\label{eq: mfhami}
\end{equation}
where 
\begin{equation}
\mathcal{E}_{0}=\mathcal{E}_{0}'+\mathcal{E}_{0}''=\sum_{\mathbf{k}}\left(A_{\mathbf{k}}-\sqrt{A_{\mathbf{k}}^{2}+\Delta^{2}}\right).
\end{equation}

\subsection{Thermodynamic potential and ground-state energy}

From the diagonalized Hamiltonian, we can straightforwardly write
down the grand thermodynamic potential, 
\begin{equation}
\Omega=-\frac{\Delta^{2}}{g}\mathcal{A}+\mathcal{E}_{0}-T\sum_{{\bf k},\eta=\pm}\ln\left(1+e^{-\frac{E_{{\bf k\eta}}}{T}}\right),
\end{equation}
where $T$ is the temperature, and the last term is the standard contribution
of non-interacting fermionic particles to the thermodynamic potential.

At $T$ = 0 K, referred to as the ground-state energy, $\Omega$ takes the following form, 
\begin{equation}
E_{G}=-\frac{\Delta^{2}}{g}\mathcal{A}+\mathcal{E}_{0}+\sum_{{\bf k},\eta=\pm}E_{{\bf k\eta}}\Theta\left(-E_{{\bf k\eta}}\right),
\end{equation}
where $\Theta(x)$ is the Heaviside step function. Since we are working on a  continuum model, it is straightforward to convert the momentum sum to an integration, and the ground-state energy density becomes:
\begin{equation}
\begin{split}\frac{E_{G}}{\mathcal{A}} & =\int\frac{d{\bf k}}{\left(2\pi\right)^{2}}\left[\sum_{\eta=\pm}E_{{\bf k\eta}}\Theta\left(-E_{{\bf k\eta}}\right)\right.\\
 & \left.+\frac{\Delta^{2}}{2\varepsilon_{{\bf k}}+\varepsilon_{B}}+A_{\mathbf{k}}-\sqrt{A_{\mathbf{k}}^{2}+\Delta^{2}}\right],
\end{split}
\label{eq: E_G}
\end{equation}
where the renormalization condition Eq. (\ref{eq: nor-cond}) is used.
At a given chemical potential, the phase boundary between SF and normal
phase is obtained by equating the ground-state energies of these two
phases: 
\begin{equation}
E_{G,\textrm{SF}}=E_{G,\textrm{N}}.
\end{equation}

\subsection{Gap equation}

During the phase transition from normal to FFLO phase, $E_{G}$ should
decrease as the onset of $\Delta$. The minimization of $E_{G}$ with
respect to $\Delta$ defines a function $S\left(t_{\textrm{AM}},{\bf q},\Delta\right)$:
\begin{equation}
-2\Delta S\left(t_{\textrm{AM}},{\bf q},\Delta\right)=\frac{\partial E_{G}}{\partial\Delta},
\end{equation}
with the explicit expression 
\begin{equation}
\frac{S\left(t_{\textrm{AM}},{\bf q},\Delta\right)}{\mathcal{A}}=\int\frac{d{\bf k}}{\left(2\pi\right)^{2}}\left[\frac{1-\sum_{\eta}\Theta\left(-E_{{\bf k\eta}}\right)}{\sum_{\eta}E_{{\bf k\eta}}}-\frac{1}{2\varepsilon_{{\bf k}}+\varepsilon_{B}}\right].\label{eq: gap-equation}
\end{equation}
Equilibrium FFLO states satisfy $S\left(t_{\textrm{AM}},{\bf q},\Delta\right)=0$,
which is also known as the gap equation. In the following, this gap
equation is utilized to find the phase boundary between normal and FFLO phase. We note that, $S(t_{\textrm{AM}},{\bf q},\Delta)$ can alternatively
be understood as the inverse vertex function (i.e., the Cooper-pair
propagator in the normal state with frequency $\omega=\Delta$) in
the ladder approximation \cite{Sheehy2015,Hu2025-2}. Therefore, the
condition $S(t_{\textrm{AM}},{\bf q},0)=0$ is precisely the well-known
Thouless criterion for the onset of the normal to SF phase
transition.

\section{Phase diagram at ${\bf q}={\bf 0}$}

In this section, we determine the phase boundaries assuming ${\bf q}=0$
pairing, which amounts to neglecting the possibility of the FFLO state.
Within this approximation, $A_{{\bf k}}=\xi_{{\bf k}}$ and $E_{{\bf k\pm}}=\sqrt{\xi_{{\bf k}}^{2}+\Delta^{2}}\pm J_{\mathbf{k}}$.
In our terminology, $E_{{\bf k\pm}}$ is non-negative. If $E_{{\bf k\pm}}>0$
for all ${\bf k}$, such a SF phase is called BCS SF. If $E_{{\bf k\pm}}=0$
at certain ${\bf k}$, then the Bogoliubov quasiparticle has a "Fermi
surface'' structure, which is also known as BFS \citep{Volovik1993}.
As discussed below, we have nodal SF with BFS protected by $\mathbb{Z}_{2}$ topological invariant. In the following, we first investigate the
BCS SF phase.

\subsection{BCS SF phase}

Since $E_{{\bf k\pm}}$ is positive, the Heaviside step function in
Eq. (\ref{eq: gap-equation}) vanishes and $S\left(t_{\textrm{AM}},{\bf 0},\Delta\right)/\mathcal{A}$
becomes 
\begin{equation}
\begin{split}\frac{S\left(t_{\textrm{AM}},{\bf 0},\Delta\right)}{\mathcal{A}} & =\int\frac{d{\bf k}}{\left(2\pi\right)^{2}}\left[\frac{1}{2\sqrt{\xi_{{\bf k}}^{2}+\Delta^{2}}}-\frac{1}{2\varepsilon_{{\bf k}}+\varepsilon_{B}}\right],\\
 & =\frac{1}{8\pi}\ln\left[\frac{\varepsilon_{B}}{\sqrt{\mu^{2}+\Delta^{2}}-\mu}\right].
\end{split}
\end{equation}
And the gap equation leads to the stationary pairing amplitude 
\begin{equation}
\Delta=\sqrt{\varepsilon_{B}\left(\varepsilon_{B}+2\mu\right)},\label{eq: upper-t_AM-1}
\end{equation}
so we must have $\mu>-\varepsilon_{B}/2$ for a stable SF phase.

The total particle number of the SF state at a fixed $\mu$ is given
by calculating $N=-\frac{E_{G,\textrm{SF}}}{\partial\mu}$, yielding,
for the total density 
\begin{equation}
n=\frac{1}{2\pi}\left(\mu+\sqrt{\mu^{2}+\Delta^{2}}\right).\label{eq: upper-t_AM-2}
\end{equation}
Inserting Eq. (\ref{eq: upper-t_AM-1}) into Eq. (\ref{eq: upper-t_AM-2})
gives results for $\Delta$ and $\mu$ for a system at a fixed total
particle number 
\begin{equation}
\begin{split}\frac{\Delta}{\varepsilon_{F}} & =\sqrt{2\frac{\varepsilon_{B}}{\varepsilon_{F}}},\\
\frac{\mu}{\varepsilon_{F}} & =1-\frac{1}{2}\frac{\varepsilon_{B}}{\varepsilon_{F}}.
\end{split}
\label{eq: upper-t_AM-3}
\end{equation}
Here, we want to emphasize that $t_{\textrm{AM}}$ does not appear
in Eq. (\ref{eq: upper-t_AM-1}), Eq. (\ref{eq: upper-t_AM-2}), and Eq.
(\ref{eq: upper-t_AM-3}). This is in agreement with the earlier result
in Ref. \citep{Sheehy2015}, where a small magnetic field does not
appear in the stationary pairing amplitude of a BCS SF phase. Therefore, both altermagnetism and Zeeman field do not effect the BCS SF phase. 

\subsection{Fixed chemical potential}

At the SF phase, the effect of NRSS disappears. The integration over
momentum and the insertion of the stationary gap-equation solution
leads to 
\begin{equation}
\frac{E_{G,\textrm{SF}}}{\mathcal{A}}=-\frac{1}{16\pi}(\varepsilon_{B}+2\mu)^{2}
\end{equation}
The effect of NRSS does appear in the normal phase. By setting $\Delta=0$
in Eq. (\ref{eq: E_G}), we find that,
\begin{equation}
\frac{E_{G,\textrm{N}}}{\mathcal{A}}=\int\frac{d{\bf k}}{\left(2\pi\right)^{2}}\left[\sum_{\eta=\pm}E_{{\bf k\eta}}\Theta\left(-E_{{\bf k\eta}}\right)+\left(\xi_{{\bf k}}-\left|\xi_{{\bf k}}\right|\right)\right],
\end{equation}
where $E_{{\bf k\eta}}=\left|\xi_{{\bf k}}\right|\pm J_{{\bf k}}$,
i.e. the energies of Bogoliubov quasiparticles reduce to the standard
single-particle excitation energies.

The integration over the second term in the above equation is straightforward:
\begin{equation}
\int\frac{d{\bf k}}{\left(2\pi\right)^{2}}\left(\xi_{{\bf k}}-\left|\xi_{{\bf k}}\right|\right)=-\frac{1}{4\pi}\mu^{2}
\end{equation}
Next we integrate over the first term. Introducing polar coordinate
${\bf k}=[k\cos\left(\phi\right),k\sin\left(\phi\right)]$ where $k$ and $\phi$ are the radial distance and polar angle respectively, and setting
$E_{{\bf k\eta}}=0$, the boundary $k_{\pm}(\phi)$ is already shown
in Fig. \ref{fig: Fig-2}(a) with the following expression: 
\begin{equation}
k_{\pm}(\phi)=\sqrt{\frac{\mu}{1\mp t_{\textrm{AM}}\sin\left(2\phi\right)}}.
\end{equation}
Here we see $t_{\textrm{AM}}<1$. $E_{{\bf k+}}$ ($E_{{\bf k-}}$)
becomes negative only when $\phi\in\left[\frac{\pi}{2},\pi\right]\cup\left[\frac{3\pi}{2},2\pi\right]$
($\left[0,\frac{\pi}{2}\right]\cup\left[\pi,\frac{3\pi}{2}\right]$),
which is marked as the cyan (blue) region in Fig. \ref{fig: Fig-2}(b).
Due to the $C_{4z}$ symmetry, it is clear that $\int\frac{d{\bf k}}{\left(2\pi\right)^{2}}E_{{\bf k+}}\Theta\left(-E_{{\bf k+}}\right)$
and $I_{1}\equiv\int\frac{d{\bf k}}{\left(2\pi\right)^{2}}E_{{\bf k-}}\Theta\left(-E_{{\bf k-}}\right)$
are the same, so only $I_{1}$ is evaluated here. What is more, the
contribution from $\phi\in\left[0,\frac{\pi}{2}\right]$ and $\phi\in\left[\pi,\frac{3\pi}{2}\right]$
are also the same, therefore, we have 
\begin{equation}
I_{1}=\frac{2}{\left(2\pi\right)^{2}}\int_{0}^{\pi/2}d\phi\int_{k_{-}(\phi)}^{k_{+}(\phi)}\left(\left|\xi_{{\bf k}}\right|-J_{{\bf k}}\right)kdk.
\end{equation}
By noticing that $\left|\xi_{{\bf k}}\right|=-\xi_{{\bf k}}$ for
$k\in\left[k_{-}(\phi),\sqrt{\mu}\right]$ and $\xi_{{\bf k}}$ for
$k\in\left[\sqrt{\mu},k_{+}(\phi)\right]$, the integration over $k$
can be calculated as 
\begin{equation}
2\int_{k_{-}(\phi)}^{k_{+}(\phi)}\left(\left|\xi_{{\bf k}}\right|-J_{{\bf k}}\right)kdk=-\frac{\left[t_{\textrm{AM}}\sin(2\phi)\right]^{2}}{1-\left[t_{\textrm{AM}}\sin(2\phi)\right]^{2}}\mu^{2},
\end{equation}
and $I_{1}$ becomes the integration over $\phi$ only: 
\begin{equation}
\begin{split}I_{1} & =-\frac{\mu^{2}}{\left(2\pi\right)^{2}}\int_{0}^{\pi/2}d\phi\left[\frac{1}{1-\left[t_{\textrm{AM}}\sin(2\phi)\right]^{2}}-1\right],\\
 & =-\frac{\mu^{2}}{\left(2\pi\right)^{2}2t_{\textrm{AM}}^{2}}\int_{0}^{2\pi}\frac{1}{\frac{2-t_{\textrm{AM}}^{2}}{t_{\textrm{AM}}^{2}}+\cos(\phi)}d\phi+\frac{\mu^{2}}{8\pi}.
\end{split}
\end{equation}
With $t_{\textrm{AM}}<1$, we have $\frac{2-t_{\textrm{AM}}^{2}}{t_{\textrm{AM}}^{2}}>1$,
the standard integration gives us 
\begin{equation}
I_{1}=\frac{\mu^{2}}{8\pi}\left(1-\frac{1}{\sqrt{1-t_{\textrm{AM}}^{2}}}\right).
\end{equation}
Therefore, we obtain 
\begin{equation}
\frac{E_{G,\textrm{N}}}{\mathcal{A}}=-\frac{1}{4\pi}\frac{\mu^{2}}{\sqrt{1-t_{\textrm{AM}}^{2}}}.\label{eq: E_G_N}
\end{equation}
It is worth noting that the area of the ellipse in Fig. \ref{fig: Fig-2}(a)
is given by $\pi ab=\pi\mu/\sqrt{1-t_{\textrm{AM}}^{2}}$, therefore,
the ground-state energy density of the normal phase is proportional
to the area of the ellipse. Such a geometric meaning can also be understood
by setting $t_{\textrm{AM}}=0$, the above ground-state energy density
becomes the well-known result $E_{G,\textrm{N}}/\mathcal{A}=-\mu^{2}/(4\pi)$.

By equating $E_{G,\textrm{SF}}$ and $E_{G,\textrm{N}}$, we have
\begin{equation}
(\varepsilon_{B}+2\mu)^{2}=\frac{\left(2\mu\right)^{2}}{\sqrt{1-t_{\textrm{AM}}^{2}}},
\end{equation}
and the critical $t_{\textrm{AM}}$ separating SF and N phase is given
by: 
\begin{equation}
t_{\textrm{AM, c}}=\sqrt{1-\left(\frac{2\mu/\varepsilon_{B}}{1+2\mu/\varepsilon_{B}}\right)^{4}}.\label{eq: t_am_c-mu}
\end{equation}
This determines the black solid curve in Fig. \ref{fig: Fig-1}(a).
When $t_{\textrm{AM}}$ is small, i.e. anisotropic spin splitting
is weak, we have BCS SF phase, even to the high chemical potential regime.
Such a feature is in accordance with the earlier result that BCS SF phase
exist at high chemical potential regime when Zeeman field is small \cite{Sheehy2015}. Large anisotropic NRSS will destabilize the BCS SF phase and drive a phase transition.

\subsection{Fixed total particle number}

After determining the phase boundary at fixed chemical potential,
our next task is to consider the case of a fixed total particle number.
By approaching the phase transition from the normal phase, $\mu=\sqrt{1-t_{\textrm{AM}}^{2}}\varepsilon_{F}$
according to Eq. (\ref{eq: E_G_N}). Then Eq. (\ref{eq: t_am_c-mu}) leads
to 
\begin{equation}
t_{\textrm{AM, c}}=\sqrt{\frac{1}{2}+\frac{\varepsilon_{B}}{2\varepsilon_{F}}-\sqrt{\frac{1}{4}-\frac{\varepsilon_{B}}{2\varepsilon_{F}}}},\label{eq: t_am_c-N}
\end{equation}
which gives the black solid curve in Fig. \ref{fig: Fig-1}(b) for $\varepsilon_{B}/\varepsilon_{F}<1/2$. When $t_{\textrm{AM}}=0$,
arbitrarily weak attraction interactions are enough to drive the system
into the BCS SF phase, as expected. The introduction of $t_{\textrm{AM}}$
will destabilize BCS SF phase and the system transits to a normal state.

\subsection{Nodal SF}

Now we study the nodal SF. Since the Bogoliubov quasiparticle energy
is 
\begin{equation}
E_{k,\pm}\left(\phi\right)=\sqrt{\xi_{{\bf k}}^{2}+\Delta^{2}}\pm t_{\textrm{AM}}k^{2}\sin\left(2\phi\right),
\end{equation}
it is highly anisotropic: along $(1,0)$ and $(0,1)$ directions,
i.e., $\phi=0$ and $\pi/2$, the pairing of the two states $\left|\mathbf{k},\uparrow\right\rangle $
and $\left|-\mathbf{k},\downarrow\right\rangle $ in Fig. \ref{fig: Fig-2}(a)
leads to an excitation energy $E_{{\bf k},\pm}=\sqrt{\xi_{{\bf k}}^{2}+\Delta^{2}}$
with minimum $\Delta$. Such a positive excitation energy follows
the spirit of BCS theory \cite{BCS1957}. Along $(-1,1)$ and $(1,1)$
direction, i.e., $\phi=\frac{3\pi}{4}(\frac{7\pi}{4})$ and $\frac{\pi}{4}(\frac{5\pi}{4})$,
$E_{\mathbf{k},+}$ and $E_{\mathbf{k},-}$ reach the same minimum
value 
\begin{equation}
E_{\mathbf{k}}=\sqrt{\xi_{{\bf k}}^{2}+\Delta^{2}}-t_{\textrm{AM}}k^{2}.
\label{eq: BdG-spectrum-min}
\end{equation}
Fig. \ref{fig: Fig-2}(c) shows the typical behavior of $E_{\mathbf{k}}$
with respect to different $t_{\textrm{AM}}$ values: $E_{\mathbf{k}}$
acquires a global minimum at $\xi_{{\bf k}}=\frac{t_{\textrm{AM}}}{\sqrt{1-t_{\textrm{AM}}^{2}}}\Delta$
with the value $E_{\mathbf{k},\textrm{min}}=\sqrt{1-t_{\textrm{AM}}^{2}}\Delta-\mu t_{\textrm{AM}}$.

BFS exists when $E_{\mathbf{k},\textrm{min}}=0$ has real positive
root ($\bar{t}_{\textrm{AM}}$) within $[0,1)$. Since $E_{\mathbf{k},\textrm{min}}=0$ leads to 
\begin{equation}
\bar{t}_{\textrm{AM}}=\sqrt{\frac{\Delta^2}{\Delta^2 + \mu^2}}, \label{eq: upper-t_AM-0}
\end{equation}
it is obvious that Eq. (\ref{eq: upper-t_AM-0}) has only one real root
within the range $[0,1)$ as shown in Fig. \ref{fig: Fig-2}(d). Therefore,
when $t_{\textrm{AM}}>\bar{t}_{\textrm{AM}}$, $E_{k,\pm}\left(\phi\right)=0$
always have solutions and we have nodal SF with the typical BFS shown
in Fig. \ref{fig: Fig-2}(e)-(f). The closed BFS separates the whole
momentum space into two regions, in the following the momentum inside
(outside) BFS is marked as $\mathbf{k}_{-}$ ($\mathbf{k}_{+}$).

By inserting Eq. (\ref{eq: upper-t_AM-1}) into Eq. (\ref{eq: upper-t_AM-0}),
we obtain the equation for determining $\bar{t}_{\textrm{AM}}$ at
fixed $\mu$:
\begin{equation}
\bar{t}_{\textrm{AM}}=\frac{\sqrt{1+2 \mu/\varepsilon_B}}{1+ \mu/\varepsilon_B}, \label{eq: upper-t_AM-0-mu}
\end{equation}
which is displayed as the black dashed line in Fig.
\ref{fig: Fig-1}(a). At the same time, by plugging Eq. (\ref{eq: upper-t_AM-3})
in Eq. (\ref{eq: upper-t_AM-0}), we obtain the equation for determining
$\bar{t}_{\textrm{AM}}$ at fixed total particle number:
\begin{equation}
\bar{t}_{\textrm{AM}}=\frac{\sqrt{2 \varepsilon_B/\varepsilon_F}}{1+ \varepsilon_B/(2\varepsilon_F)}. \label{eq: upper-t_AM-0-N}
\end{equation}
this curve is above the black line in Fig. \ref{fig: Fig-1}(b). Therefore, the nodal SF phase exists in fixed chemical potential case as $\bar{t}_{\textrm{AM}}<t_{\textrm{AM, c}}$, but vanishes in fixed total particle number
cases. 

\subsection{Topological BFS}

The fact that the BFS shown in Fig. \ref{fig: Fig-2}(e)-(f) shares
the same dimension as the normal state Fermi surface indicates it
has a topological origin. To reveal the topological invariant, here
we hinges on the Pfaffian Pf($\mathbf{k}$) of Eq. (\ref{eq: HBdG}).
Since Eq. (\ref{eq: HBdG}) preserves both charge conjuration and parity
symmetry, as suggested by Agterberg and co-workers \citep{Agterberg2017},
we can always find a suitable basis to transform Eq. (\ref{eq: HBdG})
into a screw form whose Pf($\mathbf{k}$) is given by 
\begin{equation}
\mathrm{Pf}(\mathbf{k})=\xi_{{\bf k}}^{2}+\Delta^{2}-\left[t_{\textrm{AM}}k^{2}\sin\left(2\phi\right)\right]^{2}.
\end{equation}
The solution of Pf$(\mathbf{k})$ = 0 is already given by Eq. (\ref{eq: upper-t_AM-0}).
For BFS shown in Fig. \ref{fig: Fig-2}(e)-(f), Pf$(\mathbf{k})$
is positive and negative for $\mathbf{k}_{+}$ and $\mathbf{k}_{-}$,
respectively. Hence we can identify $(-1)^{l}=\mathrm{sgn}[\mathrm{Pf}(\mathbf{k}_{-})\mathrm{Pf}(\mathbf{k}_{+})]$
as the $\mathbb{Z}_{2}$ invariant \cite{Zhao2016}.

Here we would like to compare TBFS realized here with existing approaches.
It is well known that BFS can not be realized in conventional 2D $s$-wave
superconductors through external magnetic field alone, as the required
magnetic field already drives the system into normal phase before BFS can
form. The replacement of the nodeless $s$-wave pairing by the nodal
$d_{x^{2}-y^{2}}$-wave pairing has shown to enable TBFS
\citep{Yang1998, Setty2020-1}. Alternatively, inducing supercurrent
in $s$-wave superconductors with high Fermi velocity can also give
rise to TBFS \citep{Zhu2021}. A further route involves multiband
superconductors that preserve inversion symmetry but breaks time-reversal
symmetry, as demonstrated in recent theoretical studies \citep{Brydon2018, Lapp2020, Setty2020-2}.
In contrast to these approaches, AM introduces a new but simple mechanism
for realizing TBFS. This scenario requires neither external magnetic
field nor spin-orbit coupling, but only a single band and $s$-wave
pairing, offering a minimal yet realistic platform for TBFS in superconducting
systems.

\section{FFLO phase boundary driven by NRSS}

Now we turn to the FFLO phase. To study the FFLO state, we assume
a continuous transition, with decreasing $t_{\textrm{AM}}$, from
the normal to the FFLO phase, occurring when the curvature of $E_{G}$
versus $\Delta$, at $\Delta=0$, becomes 0 for ${\bf q}\neq(0,0)$.
This equals to $S(t_{\textrm{AM}},{\bf q},0)=0$ with 
\begin{equation}
\frac{S(t_{\textrm{AM}},{\bf q},0)}{\mathcal{A}}=\int\frac{d{\bf k}}{\left(2\pi\right)^{2}}\left[\frac{N({\bf k})}{2A_{{\bf k}}}-\frac{1}{2\varepsilon_{{\bf k}}+\varepsilon_{B}}\right]\label{eq: gap-function-2}
\end{equation}
where we have defined the form factor $N({\bf k})=1-\Theta\left(-\left(A_{{\bf k}}+B_{{\bf k}}\right)\right)-\Theta\left(-\left(A_{{\bf k}}-B_{{\bf k}}\right)\right)$.
It is clear that $N({\bf k})$ exhibits discontinuities whenever an
argument of one of the Heaviside functions vanishes. 

We write $S(t_{\textrm{AM}},{\bf q},0)/\mathcal{A}$ as the sum of
the following three integrals, 
\begin{equation}
\begin{split}I_{2} & =\int\frac{d{\bf k}}{\left(2\pi\right)^{2}}\left[\frac{1}{2A_{{\bf k}}}-\frac{1}{2\varepsilon_{{\bf k}}+\varepsilon_{B}}\right],\\
I_{3} & =-\int\frac{d{\bf k}}{\left(2\pi\right)^{2}}\frac{\Theta\left[-\left(A_{{\bf k}}+B_{{\bf k}}\right)\right]}{2A_{{\bf k}}},\\
I_{4} & =-\int\frac{d{\bf k}}{\left(2\pi\right)^{2}}\frac{\Theta\left[-\left(A_{{\bf k}}-B_{{\bf k}}\right)\right]}{2A_{{\bf k}}}.
\end{split}
\end{equation}
Among them, the evaluation of $I_{2}$ is straightforward: 
\begin{equation}
I_{2}=\frac{1}{8\pi}\left(\ln\frac{\varepsilon_{B}}{2}-\ln\left|\left(1-t_{\textrm{AM}}^{2}\right)\left(\frac{{\bf q}}{2}\right)^{2}-\mu\right|\right)
\end{equation}
The evaluation of $I_{3}$ and $I_{4}$ is highly nontrivial as the
region of integration ($A_{{\bf k}}\pm B_{{\bf k}}<0$) is an ellipse
while the integrand ($1/A_{{\bf k}}$) is circular. By coordinate
transformation, $I_{3}$ and $I_{4}$ can be expressed as, 
\begin{equation}
\begin{split}I_{3} & =-\frac{ab}{2\left(2\pi\right)^{2}}\int_{0}^{2\pi}\int_{0}^{1}\frac{kdkd\phi}{f_{2}^{+}(\phi)k^{2}+f_{1}^{+}(\phi)k+f_{0}^{+}},\\
I_{4} & =-\frac{ab}{2\left(2\pi\right)^{2}}\int_{0}^{2\pi}\int_{0}^{1}\frac{kdkd\phi}{f_{2}^{-}(\phi)k^{2}+f_{1}^{-}(\phi)k+f_{0}^{-}},
\end{split}
\label{eq: integral-I3-I4}
\end{equation}
where the coefficients are 
\begin{equation}
\begin{split}f_{2}^{\pm}(\phi) & =a^{2}\cos^{2}\left(\phi\right)+b^{2}\sin^{2}\left(\phi\right),\\
f_{1}^{+}(\phi) & =\frac{(1+t_{\textrm{AM}})(q_{y}-q_{x})}{\sqrt{2}}a\cos\left(\phi\right)\\
 & +\frac{(-1+t_{\textrm{AM}})(q_{y}+q_{x})}{\sqrt{2}}b\sin\left(\phi\right),\\
f_{1}^{-}(\phi) & =\frac{(1+t_{\textrm{AM}})(q_{y}+q_{x})}{\sqrt{2}}a\cos\left(\phi\right)\\
 & +\frac{(1-t_{\textrm{AM}})(q_{y}-q_{x})}{\sqrt{2}}b\sin\left(\phi\right),\\
f_{0}^{\pm} & =\frac{{\bf q}^{2}}{2}\pm t_{\textrm{AM}}q_{x}q_{y}-\mu.
\end{split}
\end{equation}

There is a \textit{duality} between $I_{3}$ and $I_{4}$. By changing
$t_{\textrm{AM}}$ to $-t_{\textrm{AM}}$ and $\phi$ to $\phi+\frac{3\pi}{2}$
simultaneously, then $a,b$ becomes $b,a$ and $\sin(\phi),\cos(\phi)$
becomes $-\cos(\phi),\sin(\phi)$, therefore $f_{2}^{+}(\phi)$ becomes
$f_{2}^{-}(\phi)$, $f_{1}^{+}(\phi)$ becomes $f_{1}^{-}(\phi)$,
and $f_{0}^{+}$ becomes $f_{0}^{-}$. Since the translation of integral
variable $\phi\rightarrow\phi+\frac{3\pi}{2}$ does not change the
integration over $\phi$, $I_{3}$ and $I_{4}$ should be connected
by $t_{\textrm{AM}}\rightarrow-t_{\textrm{AM}}$. Such a fact simplifies
the integrations. For example, we can firstly evaluate $I_{4}$ by
contour integration, then changing $t_{\textrm{AM}}$ to $-t_{\textrm{AM}}$ to obtain $I_{3}$. 

For $d_{xy}$-wave altermagnetic spin splitting, there are two high-symmetric directions for the FF momentum $\mathbf{q}$: along
the $\left(1,0\right)$ direction (or $\left(0,1\right)$), and along
the $\left(1,1\right)$ direction (or equivalently $\left(1,-1\right)$)
according to the $C_{4z}$ rotational symmetry. The analytic evaluation of $I_3$ and $I_4$ is highly nontrivial; the calculation for $\bf{q}$ along $\left(1,1\right)$ direction in presented in  Appendix A. By contrast, the corresponding integrals for $\bf{q}$ along $\left(1,0\right)$ direction are considerably more challenging, and we have not found a closed-form expression. As discussed in Appendix
B, the angular dependence of the peak value of $S(t_{\textrm{AM}},{\bf q},0)/\mathcal{A}$
as a function of $q$ turns out to be weak. We thus anticipate that
the analysis along the $\left(1,1\right)$ direction will provide
a reasonably accurate phase boundary for the onset of the FFLO state.

\begin{figure}[t]
\begin{centering}
\includegraphics[width=0.5\textwidth]{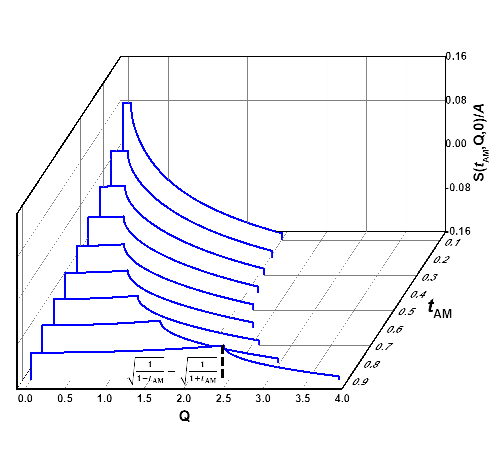}
\par\end{centering}
\caption{Plot of $S(t_{\textrm{AM}},Q,0)/\mathcal{A}$ (see Eq. (\ref{eq: SQ}) for definition) at different $t_{\textrm{AM}}$ values, here $\varepsilon_{B}=0.05$ and $\mu=1$.}
\label{fig: Fig-3}
\end{figure}

\subsection{Fixed chemical potential}

With $I_{2}$, $I_{3}$, and $I_{4}$ evaluated, introducing the dimensionless momentum $Q \equiv q/\sqrt{\mu}$ and $Q_{c} \equiv \sqrt{\frac{1}{1-t_{\textrm{AM}}}}-\sqrt{\frac{1}{1+t_{\textrm{AM}}}}$,  $S(t_{\textrm{AM}},Q,0)/\mathcal{A}$
now is calculated as: 
\begin{widetext}
\begin{equation}
\frac{S(t_{\textrm{AM}},Q,0)}{\mathcal{A}}=\frac{1}{4\pi}\begin{cases}
\ln\sqrt{\frac{2\varepsilon_{B}}{\mu}}-\ln\left|\frac{a-b}{a+b}\right|-\ln\left|\sqrt{4-(1-t_{\textrm{AM}}^{2})Q^{2}}\right|,Q\in(0,Q_c]\\
\ln\sqrt{\frac{2\varepsilon_{B}}{\mu}}-\ln\left|\frac{a-b}{a+b}\right|+\ln\left|\frac{2t_{\textrm{AM}}}{\sqrt{C(1)}+(1-t_{\textrm{AM}}^{2})+\sqrt{2D_{+}(1)}}\right|,Q\in(Q_c,\sqrt{\frac{1}{1-t_{\textrm{AM}}}}+\sqrt{\frac{1}{1+t_{\textrm{AM}}}}]
\end{cases}\label{eq: SQ}
\end{equation}
\end{widetext}
where $C(1) = (1 -t^2_{AM})^2 Q^2 + 8t^2_{AM}$ and $D_+(1) = (1-t^4_{AM})Q^2 + (1-t^2_{AM})Q\sqrt{C(1)}-4t^2_{AM}$.

Fig. \ref{fig: Fig-3} plot $S(t_{\textrm{AM}},Q,0)/\mathcal{A}$
for $\varepsilon_{B}=0.05$ and $\mu=1$. From which, it is clear that  $S(t_{\textrm{AM}},Q,0)$
increases slowly for $Q < Q_c$  and decreases rapidly for $Q > Q_c$, therefore $Q_{c}$
is a kink for $S(t_{\textrm{AM}},Q,0)/\mathcal{A}$ which reaches
the maximum value: 
\begin{equation}
\left[\frac{S(t_{\textrm{AM}},Q,0)}{\mathcal{A}}\right]_{\max}=\frac{1}{4\pi}\ln\left|\frac{\sqrt{2\varepsilon_{B}/\mu}}{\sqrt{1+t_{\textrm{AM}}}-\sqrt{1-t_{\textrm{AM}}}}\right|.
\end{equation}

It is well known that the normal phase is stable at $S(t_{\textrm{AM}},Q,0)/\mathcal{A}<0$,
therefore we see that the FFLO phase emerges at $Q=Q_{c}$
and $S(t_{\textrm{AM}},Q,0)/\mathcal{A}=0$, the simultaneous solution
of which leads to the critical $t_{\textrm{AM}}$ for FFLO phase:
\begin{equation}
t_{\textrm{AM, FFLO}}=\sqrt{1-\left(1-\frac{\varepsilon_{B}}{\mu}\right){}^{2}},\frac{\mu}{\varepsilon_{B}}\geq1\label{eq: t_AM-FFLO-mu}
\end{equation}

In Fig. \ref{fig: Fig-1}(a), we plot the FFLO phase boundary $t_{\textrm{AM, FFLO}}$
as the red curve. When $\mu<\varepsilon_{B}$, $S(t_{\textrm{AM}},Q_{c},0)=0$
has no solution and the system remains in the normal phase. For $\mu\geq\varepsilon_{B}$
the FFLO phase emerges from the normal phase side of the phase diagram,
in agreement with our assumption that the decrease of $t_{\textrm{AM}}$
leads to a continuous normal-FFLO phase transition. From Fig. \ref{fig: Fig-1}(a),
$t_{\textrm{AM}}$ has the largest window for observing FFLO phase
when $\mu=\varepsilon_{B}$. After that, the window gradually shrinks.

In Soto-Garrido \textit{et al.} \citep{Soto-Garrido2014}, the integration
over $\bf{k}$ is carried out within a shell around the Fermi surface,
and such a finite range of integration leads to a flat $S(t_{\textrm{AM}},Q,0)/\mathcal{A}$. Based on such a result, Soto-Garrido et al. \citep{Soto-Garrido2014} claimed the absence of a FFLO phase. Our results reveal the importance of full $\mathbf{k}$-integration in restoring the FFLO phase.

\subsection{Fixed total particle number}

The upper red curve, in Fig. \ref{fig: Fig-1}(b), is the location
of the continuous phase transition at a fixed total particle number.
To achieve this, we utilize $\mu=\sqrt{1-t_{\textrm{AM}}^{2}}\varepsilon_{F}$
again to obtain 
\begin{equation}
t_{\textrm{AM, FFLO}}=\sqrt{\frac{1}{2}+\frac{\varepsilon_{B}}{\varepsilon_{F}}-\sqrt{\frac{1}{4}-\frac{\varepsilon_{B}}{\varepsilon_{F}}}},\frac{\varepsilon_{B}}{\varepsilon_{F}}\leq\frac{1}{4}\label{eq: t_AM-FFLO-N}
\end{equation}
This is shown as the upper red curve in Fig. \ref{fig: Fig-1}(b).
It is clear that the larger $\varepsilon_{B}/\varepsilon_{F}$ is,
the wider the $t_{\textrm{AM}}$ window for observing FFLO phase.
Such a feature greatly facilitates the preparation of FFLO states
in electronic superconductors.

\begin{figure*}[t]
\begin{centering}
\includegraphics[width=1\textwidth]{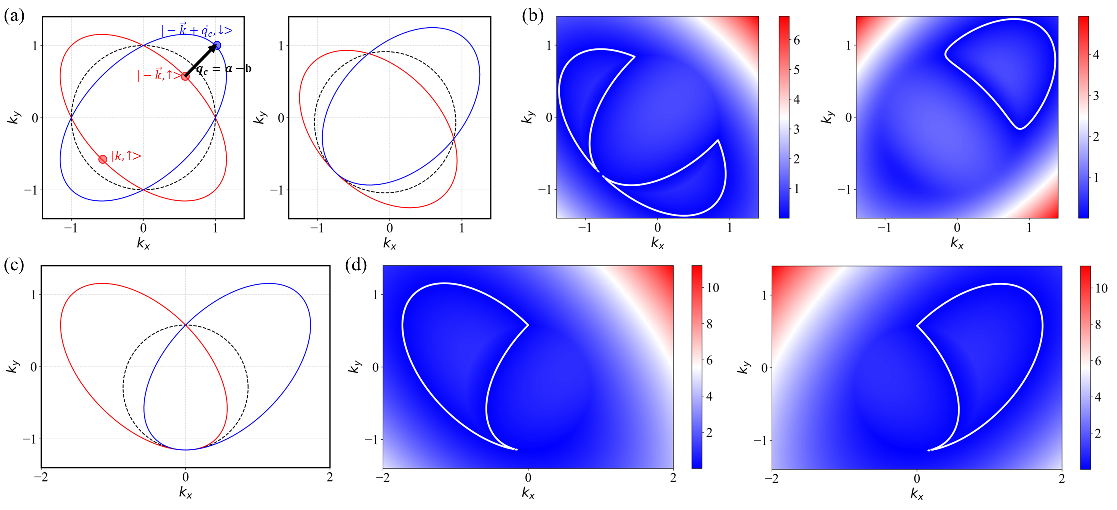} 
\par\end{centering}
\caption{(a) Illustration of Fermi surface nesting for $\mathbf{q}$ along (1, 1) direction;
(b) Bogoliubov Fermi surfaces for $\mathbf{q}=(q_{c}/\sqrt{2},q_{c}/\sqrt{2})$, here $q_c =  a - b$.
(c) Fermi surface nesting for $\mathbf{q}$ along (1, 0) direction. 
(d) Bogoliubov Fermi surfaces
for $\mathbf{q}=(q_{c},0)$, here $q_c$ is given by Eq. (\ref{eq: q_c-10}). 
$\Delta=0$, $\mu=1$, and $t_{AM}=0.5$ are chosen for plotting Bogoliubov Fermi surfaces in (b) and (d).}
\label{fig: Fig-4} 
\end{figure*}

\subsection{Geometric interpretation}

In previous section, we have shown that, for $\Delta=0$, $S(t_{\textrm{AM}},Q,0)/\mathcal{A}$ exhibits a nonanalytic dependence on $Q$ according to the contour integration of complex analysis. The nonanalyticity occurs at $Q_{c}=\sqrt{\frac{1}{1-t_{\textrm{AM}}}}-\sqrt{\frac{1}{1+t_{\textrm{AM}}}}$
which corresponds to $q_{c}=a-b$. Such an expression of $q_c$ is in agreement with Zhang \textit{et al.} \citep{ZhangS2024} where the high chemical potential approximation is made. Such a critical momentum admits a clearly geometrical interpretation as  shown in Fig. \ref{fig: Fig-4}(a),
$q_{c}$ is nothing but the momentum shift to make two of the four crossings
between the two Fermi ellipses ($E_{{\bf k}\pm}=A({\bf k})\pm B({\bf k})=0$)
and the black dashed circle ($A({\bf k})=0$) become one tangent
point. 

We note that the black dashed circle in Fig. \ref{fig: Fig-4}(a)
is where the denominator of the first term in Eq. (\ref{eq: gap-function-2})
vanishes. i.e. it represents gapless excitations. Therefore the onset
of FFLO phase here is accompanied by the decrease of gapless excitations.
Such a fact is in huge contrast with FFLO phase driven by magnetic
field, where FFLO phase emerges when the crossing point increase from
0 to 1.

The physical meaning of $q_{c}$ can also be understood through the
robustness of the BFS. Since the BFS is protected by the topological
invariant $l$, it remains stable until a topological phase transition
occurs. This transition is precisely what happens at $q_{c}$: as
shown in Fig. \ref{fig: Fig-4}(b), the number of BFS decreases from
4 to 3 at this critical momentum.

Motivated by the above analysis, we may determine $q_{c}$ when $\mathbf{q}$
is along the $(1,0)$ (or equivalently $(0,1)$) direction, where
obtaining a closed form of $S(t_{\textrm{AM}},Q,0)/\mathcal{A}$ is
difficult. Fig. \ref{fig: Fig-4}(c) demonstrates the condition when
the number of crossings decreases from 4 to 2, and the corresponding
$q_{c}$ is 
\begin{equation}
q_{c}=\frac{\sqrt{2}(a^{2}-b^{2})}{\sqrt{a^{2}+b^{2}}}=2\sqrt{\frac{\mu t_{\mathrm{AM}}^{2}}{1-t_{\mathrm{AM}}^{2}}}.\label{eq: q_c-10}
\end{equation}
The obtained BFS is shown in Fig. \ref{fig: Fig-4}(d), where the number of BFS decreases from 4 to 2. 
As elaborated in Appendix B, our numerical simulation verifies that
Eq. (\ref{eq: q_c-10}) indeed gives a good estimate of the actual $q_{c}$
when $\mathbf{q}$ is along the $(1,0)$ direction.

\section{CONCLUSIONS AND OUTLOOKS}

In summary, we have analytically investigated the ground-state phase
diagram of 2D spin-1/2 Fermi gases exhibiting $d$-wave altermagnetic
spin splitting in the presence of $s$-wave pairing, within the framework
of BdG mean-field theory. Under both fixed chemical potential and
fixed total particle number conditions, our phase diagram illustrated in Fig. \ref{fig: Fig-1} definitely
confirms the existence of FFLO states under $s$-wave pairing, thereby
resolving previous contradictory conclusions in the literatures. In
addition, we provide a geometric interpretation of the onset $\mathbf{q}_{c}$ of FFLO states, offering intuitive insight into its origin. Quite
unexpectedly, we also discover a nodal SF, characterized by the TBFS at fixed chemical potential.
Given the simplicity of the system - a single band, without external
magnetic field, and under nodeless $s$-wave pairing - the presence
of TBFS is both surprising and compelling.

Fig. \ref{fig: Fig-1} illustrates the fundamental relationship between
anisotropic NRSS and attractive pairing interaction. When the attractive pairing interaction
is strong, BCS SF phase is expected with a zero center-of-mass momentum and the Fermi surfaces are fully gapped. Meanwhile, for a very large anisotropic NRSS, the metallic
normal phase is anticipated. Nodal SF and FFLO occur when anisotropic
NRSS and attractive pairing are both at intermediate strength. All
phases can be realized in either ultracold gases or electronic superconductors
by choosing an appropriate $t_{\textrm{AM}}$. In the context of ultracold
gases, a large $t_{\textrm{AM}}$ is required as shown in Fig. \ref{fig: Fig-1}(a),
so we can access to all four phases via tuning the two-body interactions
$\varepsilon_{B}$ through Feshbach resonance \cite{Chin2010}. For
electronic superconductors, a small $t_{\textrm{AM}}$ suffices, as
shown in Fig. \ref{fig: Fig-1}(b). By increasing the Fermi energy
$\varepsilon_{F}$ via techniques like electrostatic gating \cite{Wu2023},
we can sequentially observe BCS SF, FFLO, and normal phase.

We conclude this section by noting several natural extensions of this
work. First, incorporating additional physical ingredients - such
as external magnetic fields, spin-orbit coupling, $d$-wave pairing,
and finite temperature - could significantly enrich the phase diagram
and reveal new emergent pairing phenomena. A detailed analysis of the  effects of disorder will be presented in a separated work. 
Second, while our analysis
is based on the simplest single-plane-wave ansatz here, higher plane-wave
FFLO ansatz such as the LO state with $\Delta(\mathbf{x})=\Delta\cos(\mathbf{q}\cdot\mathbf{x})$
are expected to be energetically more favorable. 
Third, our analysis relies on the mean-field framework, which provides qualitatively robust results. Quantitative refinements, with the inclusion of quantum pair fluctuations \citep{He2015} will be pursued in future work. 
Finally, the topological
BFS uncovered in this work may provide a fertile platform for realizing
Majorana bound states, opening the door to potential applications
in topological quantum computation.
\begin{acknowledgments}
We thank Jia Wang for helpful discussions. This research was supported
by the Australian Research Council's (ARC) Discovery Program, Grants
Nos. DP240101590 (H.H.) and DP240100248 (X.-J.L.). 
\end{acknowledgments}

\appendix

\section{Integration of $I_4$}

In this section, we demonstrate the evaluate of the integral $I_4$. There are two integral variables $k$ and $\phi$ in $I_{4}$. We can firstly integrate $k$ and then $\phi$ as the case in evaluating $I_1$ and $I_2$. Or we can integrate $\phi$ first and then $k$. It is found that the second choice is more convenient and is adopted here.

\begin{figure*}[t]
\begin{centering}
\includegraphics[width=1\textwidth]{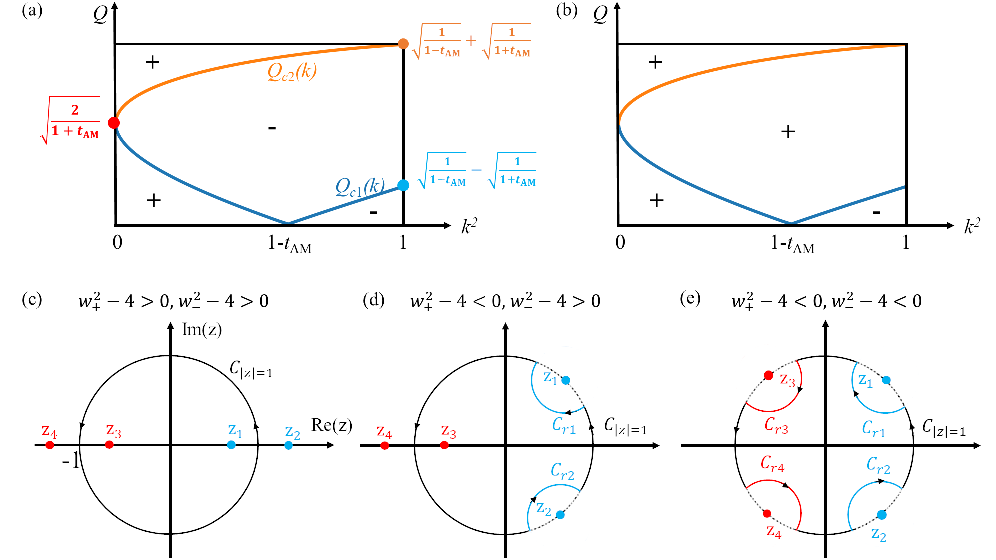} 
\par\end{centering}
\caption{(a)-(b) Sign distribution of $w_{+}^{2}-4$ and $w_{-}^{2}-4$ (see Eq. (\ref{eq: wpm-discriminant}) for definition) on the ($k, Q$) domain. The two curves $Q_{c1}(k)$ and $Q_{c1}(k)$ are given by Eq. (\ref{eq: critical-Q}). 
(c) When $w_{+}^{2}-4>0$ and $w_{-}^{2}-4>0$, $z_{1}$ and $z_{3}$ are within the unit circle while $z_{2}$ and $z_{4}$ are outside
the unit circle. At $w_{+}^{2}-4=0$, $z_{1}$ and $z_{2}$ coincide
with each other at $z=1$. After that, $w_{+}^{2}-4<0$ and $w_{-}^{2}-4>0$,
we have (d) where $z_{1}$ and $z_{2}$ are on the unit circle, while
$z_{3}$ ($z_{4}$) is inside (outside) the unit circle. When $w_{-}^{2}-4=0$,
$z_{3}$ and $z_{4}$ coincide with each other at $z=-1$. After that,
$w_{+}^{2}-4<0$ and $w_{-}^{2}-4<0$, we have (e) where all four
poles are on the unit circle. In (d)-(e), small arcs (blue and red
solid line) are added to the unit circle to form a closed contour
(solid line) to remove the poles on the unit circle. }
\label{fig: Fig-5} 
\end{figure*}

\subsection{Contour integration of $I_{4}$}

Since ${\bf q}$ is along the
$(1,1)$ direction, we have $(q_{x},q_{y})=(\frac{q}{\sqrt{2}},\frac{q}{\sqrt{2}})$
($q>0$ for simplicity) and the denominator of the integrand of $I_{4}$
is, 
\begin{equation}
\begin{split}D_{I_{4}} & =a^{2}k^{2}\cos^{2}(\phi)+b^{2}k^{2}\sin^{2}(\phi)+a(1+t_{\textrm{AM}})kq\cos(\phi)\\
 & +\frac{1+t_{\textrm{AM}}}{2}q^{2}-\mu.
\end{split}
\end{equation}
Applying Euler's formula $\cos(\phi)\rightarrow\frac{1}{2}(z+\frac{1}{z})$,
$D_{I_{4}}$ becomes 
\begin{equation}
D_{I_{4}}=\frac{(a^{2}-b^{2})k^{2}}{4}\left(z^{2}+\alpha z+\beta+\alpha z^{-1}+z^{-2}\right),
\end{equation}
where 
\begin{equation}
\begin{split}\alpha & =\frac{(1+t_{\textrm{AM}})^{2}\sqrt{1-t_{\textrm{AM}}}}{t_{\textrm{AM}}}\frac{Q}{k},\\
\beta & =\frac{2k^{2}+(1+t_{\textrm{AM}})^{2}(1-t_{\textrm{AM}})Q^{2}-2(1-t_{\textrm{AM}}^{2})}{t_{\textrm{AM}}k^{2}},
\end{split}
\end{equation}
and $Q\equiv\frac{q}{\sqrt{\mu}}$ is the dimensionless momentum.
The integral over $\phi$ now becomes a contour integral over the
unit circle: 
\begin{equation}
\begin{split}I_{4} & =-\frac{1}{(2\pi)^{2}}\frac{2ab}{a^{2}-b^{2}}\int_{0}^{1}\frac{dk}{k}\oint_{|z|=1}f_{4}(z)dz,\\
f_{4}(z) & \equiv\frac{iz}{z^{4}+\alpha z^{3}+\beta z^{2}+\alpha z+1}.
\end{split}
\end{equation}
The four poles (denoted as $z_{1/2/3/4}$) of $f_{4}(z)$ is determined
by the equation $z^{2}+\alpha z+\beta+\alpha z^{-1}+z^{-2}=0$. Such
an equation exhibits the reciprocal symmetry: if $z$ is a solution,
so is $\frac{1}{z}$. Therefore, when considering relative position
of $z_{1/2/3/4}$ with respect to the unit circle, there are three
situations: 1) all $z_{1/2/3/4}$ are on the unit circle; 2) two poles
are on the unit circle, one inside and one outside the unit circle.
Here we always choose $z_{1}$ on the unit circle (so $z_{2}=\frac{1}{z_{1}}$
is also on the unit circle) and $z_{3}$ inside the unit circle (so
$z_{4}=\frac{1}{z_{3}}$ is outside the unit circle). 3) two poles
are inside the unit circle and two outside the unit circle. We choose
$z_{1}$ and $z_{3}$ inside the unit circle, then $z_{2}=\frac{1}{z_{1}}$
and $z_{4}=\frac{1}{z_{3}}$ are outside the unit circle.

\subsection{Positions of the four poles}

Defining $w=z+\frac{1}{z}$, then $w^{2}+\alpha w+\beta-2=0$. The
discriminant is $\mathcal{D}=(1-t_{\textrm{AM}})C(k)/(t_{\textrm{AM}}^{2}k^{2})$
with 
\begin{equation}
C(k)\equiv\left(1-t_{\textrm{AM}}^{2}\right)^{2}Q^{2}+8t_{\textrm{AM}}\left(1+t_{\textrm{AM}}-k^{2}\right).
\end{equation}
With $k\in[0,1]$ and $t_{\textrm{AM}}\in(0,1)$, we have $C(k)>0$
and $\mathcal{D}>0$. Then 
\begin{equation}
w_{\pm}=\frac{\sqrt{1-t_{\textrm{AM}}}}{2t_{\textrm{AM}}k}\left[-\left(1+t_{\textrm{AM}}\right)^{2}Q\pm\sqrt{C(k)}\right],
\end{equation}
so both $w_{+}$ and $w_{-}$ are purely real.

Suppose $z_{1/2}$ ($z_{3/4}$) corresponds to the solution $z=\frac{w_{+}\pm\sqrt{w_{+}^{2}-4}}{2}$
($z=\frac{w_{-}\pm\sqrt{w_{-}^{2}-4}}{2}$). Since $w_{+}$ is purely
real, if $w_{+}^{2}-4>0$, then we have $z_{1}=\frac{w_{+}-\sqrt{w_{+}^{2}-4}}{2}$
inside the unit circle and $z_{2}=\frac{w_{+}+\sqrt{w_{+}^{2}-4}}{2}$
outside the unit circle. Nevertheless, if $w_{+}^{2}-4<0$, then we
have $z_{1/2}=\frac{w_{+}\pm i\sqrt{-w_{+}^{2}+4}}{2}$ both on the
unit circle. The same analysis applies for $w_{-}$ and $z_{3/4}$.

The two discriminants are calculated as 
\begin{equation}
w_{\pm}^{2}-4=(1+t_{\textrm{AM}})D_{\mp}(k)/(2t_{\textrm{AM}}^{2}k^{2}),
\label{eq: wpm-discriminant}
\end{equation}
with 
\begin{equation}
\begin{split}D_{\pm}(k) & \equiv\left(1-t_{\textrm{AM}}^{4}\right)Q^{2}\pm\left(1-t_{\textrm{AM}}^{2}\right)Q\sqrt{C(k)}\\
 & +4t_{\textrm{AM}}\left(1-t_{\textrm{AM}}-k^{2}\right).
\end{split}
\end{equation}
If $k^{2}<1-t_{\textrm{AM}}$, then $D_{+}(k)>0$ and $w_{-}^{2}-4>0$.
For the left cases, the condition $D_{\pm}(k)=0$ gives two critical
$Q$ curves: 
\begin{equation}
\begin{split}Q_{c1}(k) & =\left|\sqrt{\frac{2-k^{2}}{1+t_{\textrm{AM}}}}-\sqrt{\frac{k^{2}}{1-t_{\textrm{AM}}}}\right|,\\
Q_{c2}(k) & =\sqrt{\frac{2-k^{2}}{1+t_{\textrm{AM}}}}+\sqrt{\frac{k^{2}}{1-t_{\textrm{AM}}}},
\end{split}
\label{eq: critical-Q}
\end{equation}
which are shown in Fig. \ref{fig: Fig-5}(a)-(b). These two critical
$Q$ values separate the $(k,Q)$ domain into different regions
according to the sign of $w_{\pm}^{2}-4$. In Fig. \ref{fig: Fig-5}(a)
(Fig. \ref{fig: Fig-5}(b)), the region marked by "+'' sign means
$w_{+}^{2}-4>0$ ($w_{-}^{2}-4>0$), while $w_{+}^{2}-4<0$ ($w_{-}^{2}-4<0$) in the region marked by "-'' sign.

At given $(k,Q)$, there are three different possibilities according
to sign of $w_{\pm}^{2}-4$: (a) both are positive, so poles $z_{1}$
and $z_{3}$ are inside the unit circle while $z_{2}$ and $z_{4}$
outside the unit circle as shown in Fig. \ref{fig: Fig-5}(c); (b)
$w_{+}^{2}-4$ is negative while $w_{-}^{2}-4$ is positive, so poles
$z_{1}$ and $z_{2}$ are on the unit circle, while $z_{3}$ inside
and $z_{4}$ outside the unit circle, as shown in Fig. \ref{fig: Fig-5}(d);
and (c) both are negative, so all four poles are on the unit circle,
as shown in Fig. \ref{fig: Fig-5}(e). Accordingly, the integration
$I_{4}(k)\equiv\oint_{|z|=1}f_{4}(z)dz$ leads to three possibilities
and in the following, we will use $I_{4}^{(i)}(k)$ ($i=1,2,3$) to
represent these different situations.

\subsection{Integrating over $\phi$}

We first evaluate $I_{4}^{(1)}(k)$ where the poles are shown in Fig.
\ref{fig: Fig-5}(c). Since $z_{1}$ and $z_{3}$ are inside the unit
circle, the Cauchy's residue theorem gives: 
\begin{equation}
I_{4}^{(1)}(k)=2\pi i\left(\textrm{Res}[f_{4};z_{1}]+\textrm{Res}[f_{4};z_{3}]\right).
\end{equation}
Since both $z_{1}$ and $z_{3}$ are simple poles, their residues
are easy to calculate: 
\begin{equation}
\textrm{Res}[f_{4};z_{1}]+\textrm{Res}[f_{4};z_{3}]=\frac{-i}{w_{+}-w_{-}}\left(\frac{1}{z_{1}-z_{2}}+\frac{1}{z_{4}-z_{3}}\right).
\end{equation}
Inserting $w_{+}-w_{-}=\frac{\sqrt{1-t_{\textrm{AM}}}}{t_{\textrm{AM}}k}\sqrt{C(k)}$,
$z_{1}-z_{2}=-\sqrt{w_{+}^{2}-4}$ and $z_{4}-z_{3}=-\sqrt{w_{-}^{2}-4}$,
we have 
\begin{equation}
I_{4}^{(1)}(k)=-\frac{2\sqrt{2}\pi t_{\textrm{AM}}^{2}k^{2}}{\sqrt{1-t_{\textrm{AM}}^{2}}\sqrt{C(k)}}\left[\frac{1}{\sqrt{D_{+}(k)}}+\frac{1}{\sqrt{D_{-}(k)}}\right].
\end{equation}

Then we evaluate $I_{4}^{(2)}(k)$ with poles shown in Fig. \ref{fig: Fig-5}(d).
To utilize the residue theorem, we construct a closed contour shown
in Fig. \ref{fig: Fig-5}(d), where the two poles $z_{1}$ and $z_{2}$
are removed by the arcs $\mathcal{C}_{r1}$ and $\mathcal{C}_{r2}$
whose radius approaches 0. To estimate the contributions from the
two arcs: 
\begin{equation}
\begin{split}\lim_{z\rightarrow z_{1}}(z-z_{1})f_{4}(z) & =\frac{-i}{(z_{1}-z_{2})(w_{+}-w_{-})},\\
\lim_{z\rightarrow z_{2}}(z-z_{2})f_{4}(z) & =\frac{-i}{(z_{2}-z_{1})(w_{+}-w_{-})}.
\end{split}
\end{equation}
According to Jordan's lemma, the contributions of the two arcs cancel
each other. Therefore: 
\begin{equation}
\begin{split}I_{4}^{(2)}(k) & =2\pi i\textrm{Res}[f_{4};z_{3}],\\
 & =-\frac{2\sqrt{2}\pi t_{\textrm{AM}}^{2}k^{2}}{\sqrt{1-t_{\textrm{AM}}^{2}}\sqrt{C(k)}}\frac{1}{\sqrt{D_{+}(k)}}.
\end{split}
\end{equation}

For all poles on the unit circle, a closed contour can be constructed
in Fig. \ref{fig: Fig-5}(e) with no singularity inside the closed
contour. For the integration over the four arcs, it is straightforward
to prove that the contributions from $z_{1}$ and $z_{2}$ are cancelled,
so are $z_{3}$ and $z_{4}$. Therefore, $I_{4}^{(3)}(k)=0$.

\subsection{Integration over $k$}

After integrating the angular component $\phi$ via contour integration,
our next task is to integrate the radial component $k$: 
\begin{equation}
I_{4}=-\frac{1}{(2\pi)^{2}}\frac{ab}{2}\int_{0}^{1}\frac{I_{4}(k)}{k}dk,
\end{equation}
where $I_{4}(k)$ can take the form $I_{4}^{(i)}(k)$ ($i=1,2,3$)
according to $(k,Q)$.

For a given $Q$, it can intersect $Q_{c1}(k)$ and $Q_{c2}(k)$ with
either two or one crossings. Therefore, there are two domains
for $Q$: (i) $Q < Q_c$;
and (ii) $Q_c < Q <1/\sqrt{1-t_{\textrm{AM}}}+1/\sqrt{1+t_{\textrm{AM}}}$.

For $Q$ belongs to domain (i), $Q$ intersects $Q_{c1}(k)$ with
two points $k_{\pm}$ whose square are given by: 
\begin{widetext}
\begin{equation}
k_{\pm}^{2} =1-t_{\textrm{AM}}+\frac{t_{\textrm{AM}}}{2}(1-t_{\textrm{AM}}^{2})Q^{2}
 \pm\frac{1-t_{\textrm{AM}}^{2}}{2}\sqrt{4Q^{2}-(1-t_{\textrm{AM}}^{2})Q^{4}}.
\end{equation}
\end{widetext}

So the integration over [0, 1] now is split into three pieces $[0, k_-]$,  $[k_-, k_+]$, and $[k_+, 1]$, where $I_4(k)$ takes the form  $I^{(1)}_4(k)$, $I^{(2)}_4(k)$, and $I^{(3)}_4(k) = 0$, respectively. Therefore $I_{4}$ is calculated as 
\begin{widetext}
\begin{equation}
I_{4} =-\frac{1}{(2\pi)^{2}}\frac{ab}{2}\left[\int_{0}^{k_{-}}\frac{I_{4}^{(1)}(k)}{k}dk+\int_{k_{-}}^{k_{+}}\frac{I_{4}^{(2)}(k)}{k}dk\right]
 =-\frac{\sqrt{2}t_{\textrm{AM}}}{2\pi}\left[\int_{0}^{k_{-}}\frac{kdk}{\sqrt{C(k)}\sqrt{D_{-}(k)}}
+\int_{0}^{k_{+}}\frac{kdk}{\sqrt{C(k)}\sqrt{D_{+}(k)}}\right].
\label{eq: small-Q}
\end{equation}
\end{widetext}
where nest square root exists in $\sqrt{D_{\pm}(k)}$. Thanks to the
additional $\sqrt{C(k)}$, such a nest square root integration can
be evaluated in a closed form ($0\leq n<m\leq1$) 
\begin{widetext}
\begin{equation}
\int_{n}^{m}\frac{kdk}{\sqrt{C(k)}\sqrt{D_{\pm}(k)}}=-\frac{\sqrt{2}}{8t_{\textrm{AM}}}\ln\left|\frac{\sqrt{C(m)}\pm(1-t_{\textrm{AM}}^{2})Q+\sqrt{2D_{\pm}(m)}}{\sqrt{C(n)}\pm(1-t_{\textrm{AM}}^{2})Q+\sqrt{2D_{\pm}(n)}}\right|.\label{eq: nested-square-root}
\end{equation}
\end{widetext}
Inserting Eq. (\ref{eq: nested-square-root}) into Eq. (\ref{eq: small-Q}),
we obtain 
\begin{equation}
I_{4}=\frac{1}{8\pi}\ln\left|\frac{a-b}{a+b}\right|.\label{eq: I4-small-Q}
\end{equation}
For $Q$ belongs to regime (ii), $Q$ intersects $Q_{c1}(k)$ and
$Q_{c2}(k)$ at $k_{-}$. The calculation is straightforward: 
\begin{widetext}
\begin{equation}
I_{4}=-\frac{\sqrt{2}t_{\textrm{AM}}}{2\pi}\left[\int_{0}^{k_{-}}\frac{kdk}{\sqrt{C(k)}\sqrt{D_{-}(k)}}+\int_{0}^{1}\frac{kdk}{\sqrt{C(k)}\sqrt{D_{+}(k)}}\right]=\frac{1}{8\pi}\left[\ln\left|\frac{a-b}{a+b}\right|+\ln\left|\frac{\sqrt{C(1)}+(1-t_{\textrm{AM}}^{2})+\sqrt{2D_{+}(1)}}{2t_{\textrm{AM}}\sqrt{4-(1-t_{\textrm{AM}}^{2})Q^{2}}}\right|\right]\label{eq: I4-large-Q}
\end{equation}
\end{widetext}

Eq. (\ref{eq: I4-small-Q}) and Eq. (\ref{eq: I4-large-Q}) constitute
the main results of this work. The most import feature of Eq. (\ref{eq: I4-small-Q})
and Eq. (\ref{eq: I4-large-Q}) is that they are invariant under $t_{\textrm{AM}}\leftrightarrow-t_{\textrm{AM}}$.
Due to the duality between $I_{4}$ and $I_{3}$, such a feature gives
$I_{3}=I_{4}$. 

\begin{figure}[t]
\begin{centering}
\includegraphics[width=0.5\textwidth]{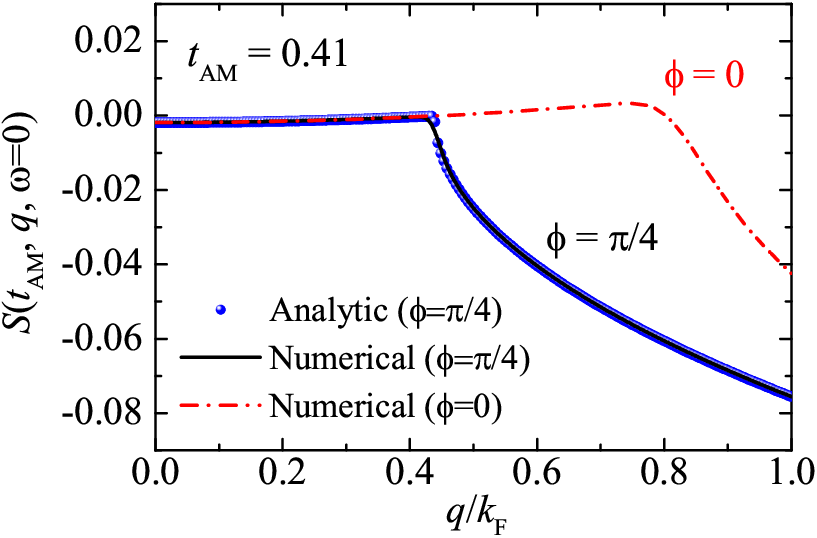}
\par\end{centering}
\caption{\label{fig: Fig6} The inverse vertex function $S(t_{\textrm{AM}},q,\omega=0)$
as a function of the center-of-mass momentum $q$, along the $x$-axis
$(1,0)$ direction (red dot-dashed line) and along the diagonal $(1,1)$
direction (black solid line and blue circles). Here, we consider an
altermagnetic coupling strength $t_{\textrm{AM}}=0.41$. The numerical
calculations are carried out at a negligible temperature $T=0.005\varepsilon_{F}/k_{B}$,
with a binding energy $\varepsilon_{B}=0.08\varepsilon_{F}$ and a
chemical potential $\mu=\sqrt{1-t_{\textrm{AM}}^{2}}\varepsilon_{F}\simeq0.912\varepsilon_{F}$.}
\end{figure}

\section{Numerical calculations of the inverse vertex function $S(t_{\textrm{AM}},\mathbf{q},\omega=0)$ }

In the inverse vertex function $S(t_{\textrm{AM}},\mathbf{q},\omega=0)$,
the two-dimensional integration over the momentum $\mathbf{k}$ could
be performed numerically. In Fig. \ref{fig: Fig6}, we present a comparison
between the numerical results and the analytic expression given in
Eq. (\ref{eq: SQ}), for the case where the center-of-mass momentum
$\mathbf{q}$ lies along the diagonal direction (i.e., the $(1,1)$
direction with $\phi=\pi/4$), at the binding energy $\varepsilon_{B}=0.08\varepsilon_{F}$.
On the scale of the figure, the two theoretical predictions appear
nearly indistinguishable, even though a small temperature $T=0.005\varepsilon_{F}/k_{B}$
has been introduced in the numerical calculation to smooth out the
sharpness of the Fermi surface.

\begin{figure}[t]
\begin{centering}
\includegraphics[width=0.5\textwidth]{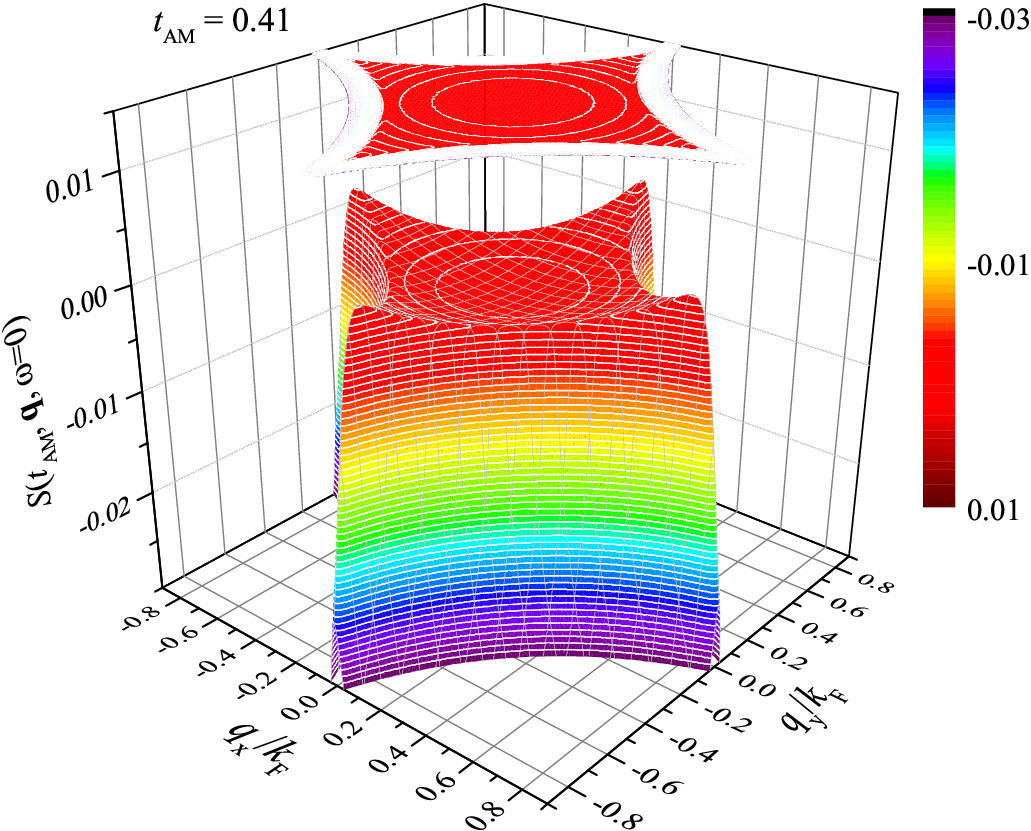} 
\par\end{centering}
\caption{\label{fig: Fig7} A three-dimensional plot of the inverse vertex function
$S(t_{\textrm{AM}},\mathbf{q},\omega=0)$ as functions of $q_{x}$
and $q_{y}$. The parameters used in the numerical calculations are
the same as in Fig. \ref{fig: Fig6}.}
\end{figure}

In Fig. \ref{fig: Fig6}, we also present the numerical result for
the case where the center-of-mass momentum $\mathbf{q}$ is along
the $x$-direction with $\phi=0$, a configuration for which an analytic
expression is not available. Interestingly, we observe that the inverse
vertex function reaches its maximum at $\phi=0$, rather than at $\phi=\pi/4$.
This trend is most clearly illustrated in the three-dimensional plot
of the inverse vertex function shown in Fig. \ref{fig: Fig7}. Despite
this angular dependence, the peak values of $S(t_{\textrm{AM}},\mathbf{q},\omega=0)$
at different $\phi$ are relatively close, implying that the critical
altermagnetic coupling strength $t_{\textrm{AM,FFLO}}$ in Eq. (\ref{eq: t_AM-FFLO-mu})
and Eq. (\ref{eq: t_AM-FFLO-N}) remains highly accurate. To verify this,
we numerically determine $t_{\textrm{AM,FFLO}}(\phi=\pi/4)=0.409$,
which is in excellent agreement with the analytic result from Eq.
(\ref{eq: t_AM-FFLO-N}), and $t_{\textrm{AM,FFLO}}(\phi=0)=0.433$. The
small discrepancy between these two critical values confirms that
it is reasonable to approximate 
\begin{align}
t_{\textrm{AM,FFLO}} & =t_{\textrm{AM,FFLO}}(\phi=0),\\
 & \simeq t_{\textrm{AM,FFLO}}(\phi=\pi/4).
\end{align}
Therefore, the red curves in the phase diagram Fig. \ref{fig: Fig-1}
are nearly unchanged, if we take into account the angular dependence
of $S(t_{\textrm{AM}},\mathbf{q},\omega=0)$ on the momentum $\mathbf{q}$.

Finally, we check the critical momentum $q_{c}$ when the center-of-mass
momentum $\mathbf{q}$ is oriented along the $(1,0)$ direction ($\phi=0$).
From the red dashed line in Fig. \ref{fig: Fig6}, we estimate $q_{c}(\phi=0)\simeq0.80k_{F}$.
This numerically determined value is in good agreement with the analytic
prediction Eq. (\ref{eq: q_c-10}) from the geometric consideration,
which yields $q_{c}(\phi=0)\simeq0.86k_{F}$ at $t_{\textrm{AM}}=0.41$. 

\end{document}